\documentclass[twocolumn,floatfix,preprintnumbers,citeautoscript]{revtex4} 
\usepackage{graphicx}
\usepackage{amsmath,amssymb,graphics}
\usepackage{bbm} 
\usepackage{accents} 
\usepackage{mathrsfs}	
\usepackage[T1]{fontenc}
\usepackage{psfrag}

\def\cl{{\cal L}}
  
\def\Bar#1{\overline{#1}}

\begin{document} \setlength{\unitlength}{1in}

\preprint{DESY 09-223}
 
\title{\Large{Leptogenesis from Quantum Interference in a Thermal Bath}}

\author{{\bf\normalsize Alexey~Anisimov,$^1$
Wilfried~Buchm\"uller,$^2$
Marco~Drewes,$^3$
Sebasti\'an~Mendizabal$^2$}\\[0.5cm]
{\it\normalsize
$^1$Fakult\"at f\"ur Physik,
Universit\"at Bielefeld, D-33615 Bielefeld, Germany}\\
{\it\normalsize
$^2$Deutsches Elektronen-Synchrotron DESY, D-22603 Hamburg, Germany}\\
{\it\normalsize
$^3$Institute de Th\'eorie des Ph\'enom\`enes Physiques
EPFL, CH-1015 Lausanne, Switzerland}\\[0.15cm]
}

\begin{abstract} 
Thermal leptogenesis explains the observed matter-antimatter asymmetry of the
universe in terms of neutrino masses, consistent with neutrino oscillation
experiments. We present a full quantum mechanical calculation of the generated
lepton asymmetry based on Kadanoff-Baym equations. Origin of the asymmetry
is the departure of the statistical propagator of the heavy Majorana neutrino
from the equilibrium propagator, together with CP violating couplings. The
lepton asymmetry is calculated directly in terms of Green's functions without
referring to `number densities'. A detailed comparison with
Boltzmann equations shows that conventional leptogenesis calculations have
an uncertainty of at least one order of magnitude. 
Particularly important is the inclusion of thermal damping rates in the
full quantum mechanical calculation.
\end{abstract}

\pacs{\dots}

\maketitle

Most theories of baryogenesis involve quantum interferences in a thermal bath
in a crucial manner \cite{kt90}. Here we consider thermal leptogenesis 
\cite{fy86} which,
in its simplest version, is dominated by the CP violating interactions of the
lightest of the heavy Majorana neutrinos, the seesaw partners of the ordinary
neutrinos. For neutrino masses inferred from neutrino oscillations, 
leptogenesis is dominated just by decays and inverse decays of the heavy
neutrinos in the thermal plasma \cite{bdp04}.

Almost all 
leptogenesis calculations are based on Boltzmann equations. This treatment has
a basic conceptual problem: the Boltzmann equations are classical equations
for the time evolution of phase space distribution functions; the involved
collision terms, however, are usually zero-temperature S-matrix elements which
involve quantum interferences. Clearly, a full quantum mechanical treatment
is necessary to understand the range of validity of the Boltzmann equations
and to determine the size of corrections \cite{bf00,bdh04}.

Various thermal corrections \cite{cnx97} have been incorporated in the context 
of Boltzmann equations, and `quantum Boltzmann equations' have been derived 
from Kadanoff-Baym equations \cite{dsr07,ghx09}. In \cite{bf00}, a solution of 
the Kadanoff-Baym equations for leptogenesis has been found to leading order 
in a derivative expansion in terms of distribution functions satisfying 
the Boltzmann equations.  

In this Letter we discuss leptogenesis directly in terms of Green's functions
which are solutions of the Kadanoff-Baym equations, thus avoiding all 
approximations necessary to arrive at Boltzmann equations. Our work is
based on \cite{abx08}, where the approach to thermal equilibrium has been 
discussed in terms of Green's functions for a toy model,
a scalar field coupled to a large thermal bath. Here we
extend this method to leptogenesis, which yields the lepton
asymmetry directly in terms of Green's functions. In the following we describe 
the main result of our work. Detailed derivations will be given in 
\cite{abx10}.
   
The interactions of $N$, the lightest of the heavy Majorana neutrinos, with the
Higgs doublet $\phi$ and lepton doublets $l_{Li}$ is described by the
lagrangian (cf.~\cite{bf00}),
\begin{align}\label{lint}
\cl &= \Bar{l}_{Li} \widetilde{\phi}\lambda^*_{i1} N + 
      N^T \lambda_{i1} C l_{Li}\phi - \frac{1}{2}M N^T C N \nonumber\\
& + \frac{1}{2}\eta_{ij} l_{Li}^T\phi\ Cl_{Lj}\phi
  + \frac{1}{2}\eta^*_{ij} \Bar{l}_{Li}\widetilde{\phi}\ C 
    \Bar{l}_{Lj}^T\widetilde{\phi}\ ;
\end{align}
here $C$ is the charge conjugation matrix, $\widetilde{\phi} = i\sigma_2 \phi$,
 and the coupling 
\begin{equation}
\eta_{ij}=\sum_{k>1}\lambda_{ik}\frac{1}{M_k}\lambda^T_{kj}
\end{equation}
is obtained by integrating out the heavy Majorana neutrinos $N_{k>1}$ with
$M_{k>1} \gg M_1 \equiv M$. We shall consider the case of small Yukawa 
couplings, 
$\lambda_{i1} \ll 1$, such that the decay width of $N$ is much smaller than
its mass. The Lagrangian (\ref{lint}) represents an effective low-energy
theory, valid for momenta up to $M_{k>1}$.

The Boltzmann equations for the time evolution of the distribution functions 
of heavy neutrinos, Higgs and lepton doublets are well known \cite{hwx09}.
In this Letter our main goal is the comparison of Boltzmann and Kadanoff-Baym
equations. We therefore focus on the CP-violating source term for the asymmetry
and ignore the washout terms and the Hubble expansion, which can be added in
a straightforward way \cite{abx10}.

For the distribution function of the heavy neutrinos one has \cite{foot1},
\begin{align}\label{bN}
\frac{\partial}{\partial t}f_N(t,\omega) =
&-\frac{2}{\omega}\int_{\bf k,q}(2\pi)^4\delta^4(k+q-p) 
\left(\lambda^{\dagger}\lambda\right)_{11} p\cdot k \nonumber\\
&\times[f_N(t,\omega)(1-f_l(k))(1+f_{\phi}(q))\nonumber\\
&\quad -f_l(k)f_{\phi}(q)(1-f_N(t,\omega))]\ ,
\end{align}
where $\omega=\sqrt{M^2+{\bf p}^2}$, $k$ and $q$ are the 
energies of $N$, $l$ and $\phi$ 
with equilibrium distribution functions $f_l$ and $f_{\phi}$, respectively; 
the averaged decay matrix element is
$|M(N(p)\rightarrow l(k)\phi(q))|^2 = 
2\left(\lambda^{\dag}\lambda\right)_{11} p\cdot k$ (cf.~\cite{bf00}). 
For the momentum integrations we use the notation
\begin{equation}
\int_{\bf p}\ldots = \int \frac{d^3 p}{(2\pi)^3 2\omega}\ldots \ .
\end{equation}
The sum of decay and inverse decay widths, which determines the rate for
the approach to equilibrium \cite{wel83}, is given by 
\begin{align}\label{gammaN}
\Gamma_{\beta}(\omega) = \left(\lambda^{\dagger}\lambda\right)_{11}
\frac{2}{\omega} \int_{\bf k,q} &(2\pi)^4\delta^4(k+q-p) \nonumber\\
& p\cdot k\ f_{l\phi}(k,q)\ ,
\end{align}
where we have introduced the function
\begin{align}
f_{l\phi}(k,q) & = f_l(k)f_{\phi}(q)+(1-f_l(k))(1+f_{\phi}(q)) \nonumber\\
& = 1 - f_l(k) + f_{\phi}(q) \ .
\end{align}
For the solution of the Boltzmann equation (\ref{bN}) with vacuum initial
condition, $f_N(0,\omega) = 0$, one easily obtains
($\Gamma_{\beta}(\omega) \equiv \Gamma$), 
\begin{equation}\label{solN}
f_N(t,\omega)=f_N^{eq}(\omega)\left(1-e^{-\Gamma t}\right),
\end{equation}
where $f_N^{eq}(\omega)=1/(e^{\beta\omega}+1)$, and $\beta=1/T$ is 
the inverse temperature.

The Boltzmann equation for the lepton distribution function is given by
\begin{align}\label{bl}
\frac{\partial}{\partial t}&f_{l}(t,k)=-\frac{1}{2k}\int_{\bf q,p} 
(2\pi)^4\delta^4(k+q-p) \nonumber\\
\times&\left[|M(l\phi\rightarrow N)|^2f_{l}(k)f_{\phi}(q)(1-f_N(t,\omega))
\right. \\
&-\left.|M(N\rightarrow l\phi)|^2f_N(t,\omega)(1-f_{l}(k))(1+f_{\phi}(q))
\right]\ , \nonumber
\end{align}
where now $\mathcal{O}(\lambda^4)$ corrections to the matrix elements have to
be kept. Using Eq.~(\ref{solN}) one obtains for the lepton asymmetry 
$f_{Li} = f_{li}-f_{\bar{l}i}$, with initial condition $f_{Li}(0,k) = 0$, 
\begin{align}\label{soll1}
f_{Li}(t,k) = - &\epsilon_{ii}\frac{1}{k}
\int_{\bf q,p}(2\pi)^4\delta^4(k+q-p)\ p\cdot k\nonumber\\
&\times f_{l\phi}(k,q) f_N^{eq}(\omega) 
\frac{1}{\Gamma}\left(1-e^{-\Gamma t}\right)\ ,
\end{align}
where
\begin{equation}
\epsilon_{ij}=\frac{3{\rm Im}\{\lambda^*_{i1}(\eta\lambda^*)_{j1}\}M}
{16\pi}\ .
\end{equation}
Summing over all flavours, the generated asymmetry is proportional to the 
familiar CP-asymmetry: $\epsilon =
\sum_i\epsilon_{ii}/\left(\lambda^{\dag}\lambda\right)_{11} 
= 3{\rm Im}(\lambda^{\dag}\eta\lambda)M/
(16\pi\left(\lambda^{\dag}\lambda\right)_{11}$) \cite{bf00}.

For later comparison with solutions of the Kadanoff-Baym equations, 
it is convenient to 
rewrite (\ref{soll1}) in the form
\begin{align}\label{soll2}
f_{Li}(t,k) = - &\epsilon_{ii}\frac{16\pi}{k}
\int_{\bf q,p,q',k'}  k\cdot k'\nonumber\\
&\times(2\pi)^4\delta^4(k+q-p)(2\pi)^4\delta^4(k'+q'-p) \nonumber\\
&\times f_{l\phi}(k,q)
f_N^{eq}(\omega)
\frac{1}{\Gamma}\left(1-e^{-\Gamma t}\right)\ .
\end{align}
Note that the integrand is now proportional to the averaged matrix element
$|M(l\phi\rightarrow \bar{l}\bar{\phi})|^2 = 2 k\cdot k' 
(\lambda^{\dagger}\lambda)_{11}/M^2$ (cf.~\cite{bf00}), which involves the 
product of the 4-vectors $k$ and $k'$.
At low temperatures, $T \ll M$, the integrand falls off like 
$e^{-\beta\omega} < e^{-\beta M}$, i.e., the generated lepton asymmetry is 
strongly suppressed.

Let us now consider spectral function and statistical propagator 
(cf.~\cite{ber04}) for the heavy Majorana neutrino,
\begin{align}
G^-_{\alpha\beta}(x_1,x_2) 
&= i\langle \{N_{\alpha}(x_1),N_{\beta}(x_2)\}\rangle\ ,\\
G^+_{\alpha\beta}(x_1,x_2)
&={1\over 2}\langle [N_{\alpha}(x_1),N_{\beta}(x_2)]\rangle\ ,
\end{align}
which satisfy the Kadanoff-Baym equations \cite{bf00,abx10} 
\begin{align}
C(i\gamma^0\partial_{t_1}-&{\bf p}\pmb{\gamma}-M)G_{\bf p}^-(t_1-t_2)=
\nonumber\\
&-\int_{t_2}^{t_1} dt'\Sigma_{\bf p}^{-}(t_1-t')G_{\bf p}^-(t'-t_2)\ ,\\
C(i\gamma^0\partial_{t_1}-&{\bf p}\pmb{\gamma}-M)G_{\bf p}^+(t_1,t_2) =
\nonumber\\ 
&+\int^{t_2}_0dt'\Sigma_{\bf p}^{+}(t_1-t')G_{\bf p}^-(t'-t_2)\nonumber\\
&-\int^{t_1}_{0}dt' \Sigma^-_{\bf p}(t_1-t')G_{\bf p}^{+}(t',t_2)\ .
\label{kb}
\end{align}
Here we have assumed spatial homogeneity and performed a Fourier transform.
The 1-loop contribution to the self-energies $\Sigma^{\pm}_{\bf p}$ is shown
in Fig.~1. For small couplings, $\lambda \ll 1$, leading to a small width
$\Gamma \ll M$, explicit solutions of the Kadanoff-Baym equations can be
found in the Breit-Wigner approximation \cite{abx10},
\begin{figure}[t] 
    \psfrag{p}{$\phi$}\psfrag{wp}{$(\omega, {\bf p})$}
        \psfrag{l}{$l$}
            \psfrag{N}{$N$}
   \centering
   \includegraphics[width=2in]{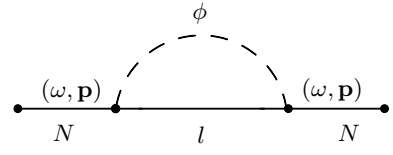} 
   \caption{1-loop contribution to the self-energies $\Sigma^{\pm}_{\bf p}$ 
of the Majorana neutrino $N$.}
   \label{fig:1loop}
\end{figure}
\begin{align}
G^-_{\textbf{p}}(y) &= \left(i\gamma_0\cos(\omega y)
+\frac{M-{\bf p}\pmb{\gamma}}{\omega}\sin(\omega y)\right)
\label{g-}\\
&\quad \times e^{-\Gamma|y|/2}C^{-1}\ , \nonumber\\
G^{+}_{ \textbf{p}}(t,y) &= 
- \left(i\gamma_0\sin(\omega y)
-\frac{M-{\bf p}\pmb{\gamma}}{\omega}
\cos(\omega y)\right) \label{g+}\\
&\quad\times
\left[\frac{\tanh\left(\frac{\beta\omega}{2}\right)}{2}
e^{-\Gamma|y|/2} + f_N^{eq}(\omega) e^{-\Gamma t}\right]C^{-1} ,
\nonumber
\end{align}
where $t=(t_1+t_2)/2$, $y=t_1-t_2$ and $\Gamma=\Gamma_{\beta}(\omega)$ 
(\ref{gammaN}). For large $t$, $G^+_{\bf p}(t,y)$ approaches
the equilibrium solution $G^{+eq}_{\bf p}(y)$, and for small temperatures, 
i.e., large $\beta$, it becomes the vacuum solution $G^{+vac}_{\bf p}(y)$. 
Note that the solution (\ref{g+}) for  $G^{+}_{\bf p}(t,y)$ satisfies the
initial condition
\begin{equation}
G^{+}_{\bf p}(0,0) = G^{+vac}_{\bf p}(0)\ ,
\end{equation}
which is the analogue of the initial condition $f_N(0,\omega) = 0$ for the
distribution function. For spectral function and statistical propagator 
of lepton and Higgs fields we shall use the free equilibrium expressions
$\hat{S}^{\pm}_{\bf k}(y)$ and $\hat{\Delta}^{\pm}_{\bf q}(y)$, respectively.
The thermal equilibrium is assumed to be established by Standard Model
interactions.

\begin{figure}[t] 
    \psfrag{p}{$\phi$}
        \psfrag{p}{$\phi$}
        \psfrag{k}{$\bf{k}$}
            \psfrag{q}{$\bf{q}$}
	\psfrag{kp}{$\bf{k}'$}
         \psfrag{qp}{$\bf{q}'$}
                \psfrag{omega}{$(\omega,\bf{p})$}
    \psfrag{t1}{$t$}
    \psfrag{t2}{$t'$}
    \psfrag{t3}{$t_1$}
    \psfrag{t4}{$t_3$}
    \psfrag{t5}{$t_2$}
        \psfrag{l}{$l$}
            \psfrag{N}{$N$}
            \psfrag{+}{$+$}
   \centering
   \includegraphics[width=3in]{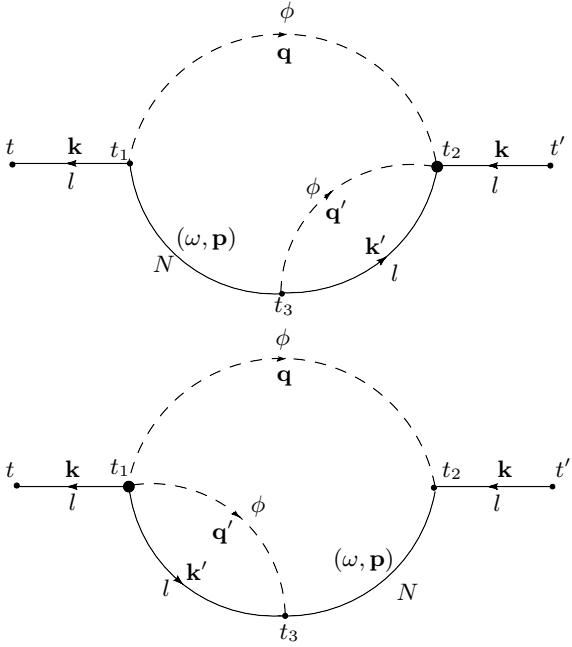} 
   \caption{2-loop contributions to the lepton self-energies 
$\Pi^{\pm}_{{\bf k}}$, which lead to non-zero lepton number densities.}
\end{figure}

We are now ready to calculate the lepton asymmetry which is generated during
the approach of the right-handed neutrino $N$ to equilibrium.
The `lepton number matrix' is obtained from the statistical propagator
of the lepton fields, 
\begin{equation}
L_{{\bf k}ij}(t,t') =   
- \mathrm{tr}[\gamma^0 S^+_{{\bf k}ij}(t,t')]\ .
\end{equation}
One easily verifies that for free fields in equilibrium, 
$L_{{\bf k}ii}|_{t=t'} = f_{li}(k)-\bar{f}_{li}(k)$, which vanishes for
zero chemical potential. 
To leading order in $\lambda$, a flavour 
non-diagonal asymmetry is generated by the 2-loop self-energies shown in Fig.~2
(cf.~\cite{bf00}). Solving the Kadanoff-Baym equation for $S^+_{\bf k}$
to first order in the self-energy $\Pi_{\bf k}^{\pm}$, one finds after some
algebra 
\begin{align}\label{nonlocal}
L_{{\bf k}ij}(t,t) = -i \int_0^{t} dt_1  &\int_0^{t}dt_2 \ 
\mathrm{tr}[\hat{S}^+_{\bf k}(t_2-t_1)\Pi^-_{{\bf k}ij}(t_1,t_2)
\nonumber\\
&- \hat{S}^-_{\bf k}(t_2-t_1)\Pi^+_{{\bf k}ij}(t_1,t_2)]\ .
\end{align}

A non-zero asymmetry is generated by the departure of $G^+_{\bf p}$ from
equilibrium. Due to the chiral couplings only a chiral projection of 
$G^+_{\bf p}$ contributes,
\begin{align}
\tilde{G}_{\bf p}(t,y) P_L &= 
P_L \left(G^+_{\bf p}(t,y) - G^{+eq}_{\bf p}(y)\right)P_L \ , \nonumber\\ 
\tilde{G}_{\bf p}(t,y) &= \frac{M}{\omega}
\cos(\omega y) f_N^{eq}(\omega) e^{-\Gamma t}\ .
\end{align}
After a lengthy calculation one obtains for the lepton number matrix
to leading order in the width $\Gamma$ \cite{abx10}:
\begin{align}\label{asymkb}
L_{{\bf k}ij}(t,t) &=  -\epsilon_{ij}\ 8\pi 
\int_{\bf q,q'} \frac{k\cdot k'}{kk'\omega} \nonumber\\
&\times\frac{\frac{1}{2}\Gamma}{((\omega-k-q)^2 +\frac{\Gamma^2}{4})
((\omega-k'-q')^2 + \frac{\Gamma^2}{4})} \nonumber\\
&\times f_{l\phi}(k,q) f_{l\phi}(k',q') f_N^{eq}(\omega)\nonumber\\
&\times\left(\cos[(k+q-k'-q')t] + e^{-\Gamma t}  \right. \\
&\left. - (\cos[(\omega-k-q)t] 
 +\cos[(\omega-k'-q')t]) e^{-\frac{\Gamma t}{2}}\right) , \nonumber
\end{align}
where $\bf{p}=\bf{q}+\bf{k} = \bf{q}'+\bf{k}'$. This expression for the lepton 
asymmetries, generated by
quantum interference in a thermal bath, is the main result of this
Letter.

The expression (\ref{asymkb}) contains a logarithmic divergence 
$\mathcal{O}(\lambda^4)$. It has to be combined with other 
divergent terms which have been neglected in (\ref{asymkb}) since they do 
not contribute to leading order in $\Gamma$. The total divergence will be 
subtracted by a counter term. The dominant finite contribution to the
integral stems from momenta $k+q \sim k'+q' \sim \omega$. In the zero-width
limit $\Gamma \rightarrow 0$, with $\Gamma t$ fixed, the integrand is
$\mathcal{O}(1/\Gamma)$. Since $\Gamma \propto \lambda^2$, this dominant
contribution to the integral is $\mathcal{O}(\lambda^2)$.

It is interesting to compare the diagonal elements of the lepton number matrix
$L_{{\bf k}ij}(t,t)$ with the distribution functions $f_{Li}(t,k)$ given in 
(\ref{soll2}). As expected, the same CP asymmetries $\epsilon_{ii}$  
appear, whereas the dependence of the integrands on time and temperature
is different. The reason for the different temperature dependence is the
fact that the matrix elements in the Boltzmann equations were calculated
at zero temperature. Hence, the factor $f_{l\phi}(k',q')$ is missing in
(\ref{soll2}). Since Boltzmann equations are local in time
whereas Kadanoff-Baym equations contain `memory effects',
the different time dependence of the asymmetries is also expected.
One consequence is that $\partial_t f_{Li}(t,k)|_{t=0}\neq 0$, whereas 
$\partial_t L_{{\bf k}ij}(t,t)|_{t=0}=0$. Particularly important are 
off-shell effects in (\ref{asymkb}), which lead to terms oscillating 
in time.    

The additional thermal correction $f_{l\phi}(k',q')$ is linear in the
distribution functions $f_{l}$ and $f_{\phi}$, in contrast to
results obtained in the first two papers in \cite{cnx97}, 
but in agreement with \cite{dsr07} and \cite{ghx09}. 
The memory and off-shell effects
found in \cite{dsr07} are qualitatively different from our result
(\ref{asymkb}).

It is instructive to consider the approximations, even if they might not be
justified, which lead from (\ref{asymkb}) to the result of Boltzmann 
equations. Neglecting off-shell effects, i.e., imposing 
$\omega = k + q = k' + q'$, 
the cosines are replaced by one; performing then 
the zero-width approximation  $\Gamma \rightarrow 0$, with $\Gamma t$ 
fixed, the integral (\ref{asymkb}) becomes  
\begin{align}\label{asymkb2}
L_{{\bf k}ij}^{os}(t,t) &= - \epsilon_{ij}\frac{16\pi}{k}
\int_{\bf q,q',p,k'} k\cdot k' \nonumber\\
&\times(2\pi)^4\delta^4(k+q-p)(2\pi)^4\delta^4(k'+q'-p)\nonumber\\
&\times f_{l\phi}(k,q) f_{l\phi}(k',q') f_N^{eq}(\omega)\nonumber\\
&\times\frac{1}{\Gamma}\left(1 - e^{-\frac{\Gamma t}{2}}\right)^2\ .
\end{align}
Except for the factor $f_{l\phi}(k',q')$, the only difference compared to
the solution (\ref{soll2}) of the Boltzmann equations is the time dependence.
It is obvious from Fig.~2 and Eq.~(\ref{nonlocal}) that in the quantum
theory the generation of the lepton asymmetry is nonlocal in time. This
leads to the square of the exponential fall-off in (\ref{asymkb2}).
On the contrary, in the Boltzmann
equations the asymmetry is generated locally in time yielding a
simple exponential behaviour. The difference can be numerically important 
at cosmologically relevant times $t_L \sim 1/\Gamma$.

The calculations leading to Eq.~(\ref{asymkb}) also demonstrate that the
result for the lepton asymmetry will be significantly modified by the 
thermal damping rates for lepton and Higgs fields in the plasma. These 
thermal widths
(cf.~\cite{amy02}) are known to be much larger than the decay width of the 
Majorana neutrino: $\Gamma_l \sim \Gamma_{\phi} \sim g^2 T \gg \lambda^2 M$
for $M \lesssim T$. For quantum interferences the thermal damping rates are 
qualitatively more important than thermal masses which, for simplicity, we
ignore in the following. Including naively thermal widths for lepton and 
Higgs fields in the self-energy $\Pi^+_{\bf k}$, one obtains instead of 
(\ref{asymkb}) to leading order in these widths 
($\Gamma_{l\phi}=\Gamma_l + \Gamma_{\phi}$)
\begin{align}\label{asymdamp}
\tilde{L}_{{\bf k}ij}(t,t) &= -\epsilon_{ij}\ 16\pi \int_{{\bf q,q'}}
\frac{k\cdot k'}{kk'\omega}\nonumber\\
&\times\frac{\frac{1}{4}\Gamma_{l\phi}\Gamma_{\phi}}  
{((\omega-k-q)^2+\frac{1}{4}\Gamma_{l\phi}^2)
((\omega-k'-q')^2+\frac{1}{4}\Gamma_{\phi}^2)}\nonumber\\
&\times f_{l\phi}(k,q) f_{l\phi}(k',q') f_N^{eq}(\omega)\nonumber\\
&\times\frac{1}{\Gamma}\left(1-e^{-\Gamma t}\right).
\end{align}
Note that the factors oscillating in time have disappeared. The thermal
widths of lepton and Higgs fields have led to a behaviour which is local 
in time. In the zero-width limit one now obtains the result 
(\ref{soll2}) of
the Boltzmann equations except for the thermal correction factor 
$f_{l\phi}(k',q')$.
We emphasize that (\ref{asymdamp}) is speculative at present, and it remains
to be seen whether it follows from a solid calculation which includes
gauge interactions in a systematic way.

We have studied the generation of a lepton asymmetry at constant temperature
as the heavy Majorana neutrino approaches thermal equilibrium. For
cosmological leptogenesis one has to calculate the lepton asymmetry in the
case of decreasing temperature, which is caused by the expansion of the 
universe. Analyses based on Boltzmann equations suggest that both asymmetries
are of comparable size for $T \sim M$ \cite{bdp04}. A more detailed discussion 
will be given in \cite{abx10}.

In this Letter we have compared lepton asymmetries calculated on the basis 
of Boltzmann
equations with those obtained from Kadanoff-Baym equations.
Our discussion illustrates, that even ignoring spectator and flavour effects
\cite{bcx99}, current leptogenesis calculations have an uncertainty of at 
least one order of magnitude. Particularly urgent is the inclusion of
gauge interactions with the thermal bath in a full quantum calculation.\\

\textbf{Acknowledgements.} 
We would like to thank D.~B\"odeker, L.~Covi and O.~Philipsen for helpful
discussions.
AA and WB acknowledge support of the German Research Foundation (DFG) in the
SFB 676 ``Particles, Strings and the Early Universe''; SM has been supported
by the German Academic Exchange Service (DAAD).

\end{document}